\documentclass[pra,superscriptaddress,showpacs,papersize=a4paper,twocolumn,floatfix]
{revtex4-1}%
\usepackage{float}
\usepackage{graphicx}
\usepackage[usenames,dvipsnames]{color}
\usepackage{amsmath}
\usepackage{bm}
\usepackage{amsfonts}
\usepackage{amssymb}
\def\be{\begin{equation}}
\def\ee{\end{equation}}

\def\Tr{\mathop{\rm Tr}}

\newcommand{\corr}[1]{\langle #1\rangle}
\renewcommand{\Re}{\mathop{\rm Re}}
\renewcommand{\Im}{\mathop{\rm Im}}
\newcommand{\eps}{\varepsilon}

\begin{document}
\title{Superconductivity in the presence of microwaves: Full phase diagram}
\begin{abstract}
We address the problem of non-equilibrium superconductivity in the presence of microwave irradiation. Using contemporary analytical methods, we refine the old Eliashberg theory and generalize it to arbitrary temperatures $T$ and frequencies $\omega$. Microwave radiation is shown to stimulate superconductivity in a bounded region in the $(\omega,T)$ plane. In particular, for $T<0.47\, T_c$ and for $\hbar\omega>3.3\, k_BT_c$ superconductivity is always suppressed by a weak \emph{ac} driving. We also study the supercurrent in the presence of microwave irradiation and establish the criterion for the critical current enhancement. Our results can be qualitatively interpreted in terms of the interplay between the kinetic (``stimulation" vs.\ ``heating") and spectral (``depairing") effects of the microwaves.
\end{abstract}

\author{K. S. Tikhonov}
\affiliation{Skolkovo Institute of Science and Technology, Skolkovo 143026, Russia}
\affiliation{L. D. Landau Institute for Theoretical Physics,
Chernogolovka 142432, Russia}

\author{M. A. Skvortsov}
\affiliation{Skolkovo Institute of Science and Technology, Skolkovo 143026, Russia}
\affiliation{L. D. Landau Institute for Theoretical Physics,
Chernogolovka 142432, Russia}

\author{T. M. Klapwijk}
\affiliation{Kavli Institute of Nanoscience, Faculty of Applied Sciences,
Delft University of Technology, 2628 CJ Delft, The Netherlands}
\affiliation{Physics Department, Moscow State University of Education, Moscow 119992, Russia}

\date{\today}

\maketitle
\section{Introduction}

The full understanding of the non-equilibrium properties of superconductors is important for both fundamental theory and applications. One of the basic phenomena in this field is the microwaves enhancement of superconductivity, known for constriction-type microbridges as the Dayem-Wyatt effect \cite{dayem1967behavior,wyatt1966microwave}. The basic form of this effect is generally observed in superconducting stripes and amounts to enhancement of the superconducting gap due to a non-equilibrium distribution of quasiparticles created by a microwave field. It was theoretically explained by Eliashberg \cite{eliashberg70,ivlev1971influence} on the basis of the dynamic Gorkov equations \cite{gorkov1969superconducting}. Since the superconducting gap $\Delta$ is not easily available directly, the influence of the microwaves on the critical pair-breaking 
current $I_c$ and the critical temperature $T_c$ can be more preferable for experimental study. Klapwijk and Mooij reported \cite{klapwijk1976microwave,klapwijk1977radiation} the observation of the enhancement of $I_c$ and, most notably, also $T_c$ of long homogeneous strips. Direct observation of the gap enhancement followed in Ref.~\onlinecite{kommers1977measurement}. This field flourished for years and the state of the art at 1980s was summarized in the review \cite{mooij1981enhancement}.

\begin{figure}[b]
\centering%
\includegraphics[width=0.47\textwidth]{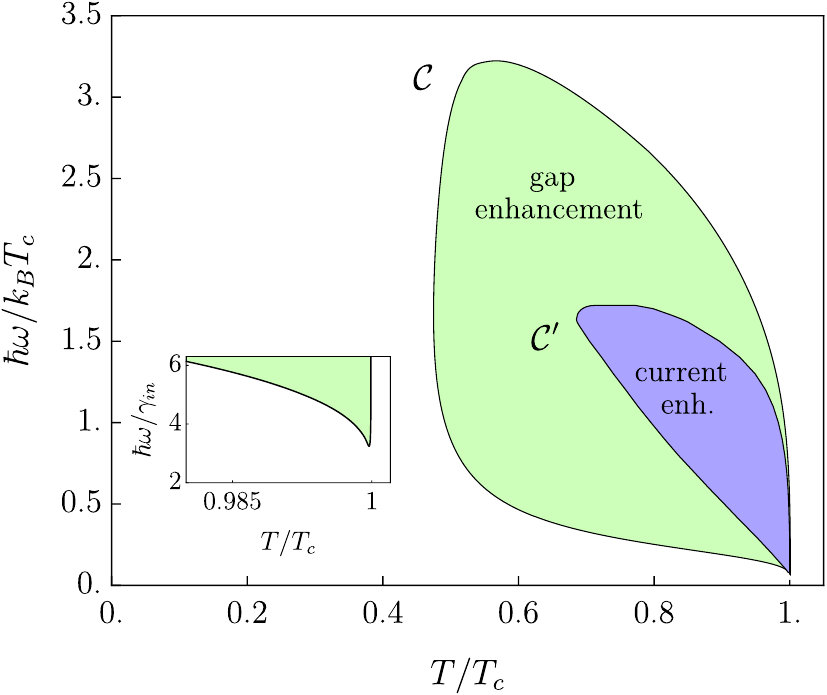}
\caption{Phase diagram of a superconductor under weak microwave driving ($\alpha\to 0$) at the frequency $\omega$ (the inelastic relaxation rate $\gamma_\text{in}/k_BT_c=0.02$). Gap enhancement is observed inside the curve $\mathcal{C}$. The region of the critical current enhancement is bounded by the curve $\mathcal{C}'$.
Inset: zoom of the gap enhancement region near $T_c$, showing the minimal frequency $\omega_\text{min,min}\approx
3.23 \, \gamma_\text{in}/\hbar$.}
\label{wmin-gap}
\end{figure}

In developing the Eliashberg theory, more accurate models of inelastic relaxation (realistic electron-phonon interaction) were introduced \cite{chang1977gap}, including an additional contribution to the enhancement by the energy-dependence of the recombination rate. The important issue of stability of the out-of-equilibrium superconducting phase was studied by Schmid and co-workers \cite{schmid1977stability,eckern1979stability}. Interestingly, although enhancement of the critical current was the first experimental manifestation of the effect, its microscopic theory was lacking for a while until the supercurrent flow in a superconductor under out-of-equilibrium conditions was evaluated in Ref.~\onlinecite{schmid1980dynamic}. Shortly, the current dependence of the superconductivity enhancement was studied in detail experimentally \cite{van1984microwave}. As one of the fundamental features of the non-equilibrium response is its strong sensitivity to inelastic processes, it is possible to use it as a direct measure of the strength of these processes. A direct proportionality between the minimum irradiation frequency required for the enhancement of the critical current and the inelastic scattering rate was used in Ref.~\onlinecite{van1984inelastic} for such a measurement. Similar ideas have been discussed theoretically for superconducting weak links \cite{schmid1980dynamic} and SNS junctions \cite{lempitskii1983stimulation,virtanen2011linear,tikhonov2015admittance}, and studied in much detail in recent experiments \cite{chiodi2011probing,dassonneville2013dissipation}.

Superconductivity enhancement in both homogeneous systems (superconducting wires and films) and hybrid structures is associated with the fact that the quasiparticle distribution function as a function of energy acquires structure at the sub-thermal scale (the superconducting gap $\Delta$ in the former case and the minigap $\epsilon_g$ in the latter case). However, while the microwave field drives quasiparticles out of equilibrium, it is not the only effect. It is indeed the leading one sufficiently close to the critical temperature, when the density of states (DOS) available for excitations is large. At lower $T$, a modification of the order parameter by the microwaves becomes more and more important. It is well known that even under equilibrium conditions, the DOS in a current-carrying superconductor is non-trivial \cite{Maki-Parks,anthore2003density}. 
As shown recently by Semenov \emph{et al}.\  \cite{semenov2016coherent}, under driving by microwaves the spectral properties of the superconducting wire are strongly modified by the field even at zero temperature and coherent excited states are formed.

In the present work, we study the spectral and kinetic response of a current-carrying superconducting wire to the microwaves. We consider a diffusive superconductor (elastic mean free path much shorter than the BCS coherence length $\xi_0$) irradiated by an \emph{ac} electromagnetic wave in the presence of a \emph{dc} supercurrent described by a constant vector potential. We assume energy relaxation to result from tunneling to a nearby equilibrium normal reservoir with an energy-independent rate $\gamma_\textrm{in}={\hbar}/{\tau_\text{in}}$. Such a model is formally equivalent to the relaxation time approximation used by Eliashberg and co-workers \cite{eliashberg70,ivlev1971influence}.  We assume a quasi-one-dimensional geometry, so that both the \emph{ac} and \emph{dc} components of the vector potential are collinear with the wire. We treat the \emph{ac} field as a perturbation but impose no constraints on the temperature $T$, frequency $\omega$, order parameter $\Delta$, \emph{dc} component of the vector potential $A_0$, and the energy relaxation rate~$\gamma_\textrm{in}$.

In the framework of the described model, our results are summarized in the phase diagram shown in Fig.~\ref{wmin-gap}. The curve $\mathcal{C}$ encircles the region in the $(\omega,T)$ plane, where relatively weak ($\alpha\to0$) electromagnetic irradiation actually enhances the superconducting gap $\Delta(T)$ with respect to its equilibrium BCS value $\Delta_0(T)$ in the absence of a supercurrent. Importantly, this region has natural bounds from the side of low temperatures (due to vanishing of the available DOS) and from the side of high frequencies (the field oscillating too fast is unable to create strong enough out-of-equilibrium population and simply heats the system). 
The curve $\mathcal{C}'$ in Fig.~\ref{wmin-gap} encloses the region where the critical current of the superconductor is enhanced by microwave irradiation. 
The region of the critical current enhancement is narrower than the region of the gap enhancement, illustrating a simple fact that it is actually \emph{harder} to enhance the superconductivity when the current is applied. This is a result of the pair-breaking effect of the supercurrent, which smoothens the singularity in the BCS DOS \cite{Maki-Parks,anthore2003density} and, hence, in the 
field-induced distribution function of quasiparticles.

The paper is structured as follows. In Sec.\ \ref{sec:eli} we discuss the main ingredients of the Eliashberg theory of superconductivity enhancement. In Sec.\ \ref{S:theory} we formulate our $\sigma$-model-based approach, valid in the whole region of parameters of the problem. Next, we describe the results in Sec.\ \ref{sec:res} and conclude in Sec.\ \ref{sec:summary}.

\section{Eliashberg theory ($T\to T_c$)}
\label{sec:eli}

The standard theory of gap enhancement pioneered by Eliashberg \cite{eliashberg70,ivlev1971influence}, elaborated in Refs.~\cite{schmid1977stability,eckern1979stability} and extended to treat the supercurrent \cite{schmid1980dynamic,van1984microwave,van1984inelastic} describes a diffusive superconductor subject to microwave irradiation in the vicinity of the critical temperature. It assumes that the absolute value of the order parameter is uniform over the sample. Then gauging out the phase of the order parameter one arrives at a zero-dimensional problem
in the field of a time-dependent vector potential
\be
\label{A}
\mathbf{A}(t) = \mathbf{A}_{0}+\mathbf{A}_{1}\cos\omega t,
\ee
where the static part $\mathbf{A}_0$ accounts for the \emph{dc} supercurrent, and $\mathbf{A}_1{\parallel}\mathbf{A}_0$.
To characterize the depairing effect of the vector potential \cite{Maki-Parks} it is convenient to introduce the energy scales (depairing rates)
\be
\label{Gamma}
\Gamma = \frac{2e^{2}D\mathbf{A}_0^2}{\hbar c^2},
\qquad
\alpha = \frac{2e^{2}D\mathbf{A}_1^2}{\hbar c^2},
\ee
where $D$ is the normal-state diffusion coefficient in the superconductor \cite{defalpha}.

The Eliashberg theory naturally generalized to the presence of a finite $\mathbf{A}_0$ provides the following GL equation for the time-averaged order parameter $\Delta$:
\begin{equation}
\frac{7\zeta(3)}{8\pi^2}\left(\frac{\Delta}{k_{B}T_{c}}\right)^2
-
\frac{T_{c}-T}{T_{c}}
+
\frac{\pi\Gamma}{4k_{B}T_{c}}
=
\alpha \mathcal{F}_\text{neq} ,
\label{GL}
\end{equation}
where the left-hand side is the usual expansion in the absence of radiation (with the last term describing depairing due to the supercurrent), while the right-hand side perturbatively accounts for the \emph{ac} component of the vector potential.
In general, expression for $\mathcal{F}_\text{neq}$ is a complicated function of $\omega$, $\Delta$, $\Gamma$ and $\gamma_\textrm{in}$ (see Sec.\ \ref{SS:our-gap-enh}).
The Eliashberg theory assumes inelastic relaxation to be the slowest process and considers the limit
\be
\gamma_\textrm{in}\ll(\hbar\omega,\Delta)\ll k_B T .
\label{Eliash-cond}
\ee
Under these conditions the function $\mathcal{F}_\text{neq}$ in the right-hand side of Eq.\ (\ref{GL}) acquires the form
\be
\mathcal{F}_\text{neq}
=
- \frac{\pi}{8k_{B}T_{c}}
+
\frac{\hbar\omega}{16\gamma_\textrm{in}k_BT_c} G\left(\frac{\Delta}{\hbar\omega}, \frac{\Gamma}{\Delta}\right) ,
\label{OmegaA}
\ee
where the first term is due to the modification of the static spectral functions (depairing), while the second term has a kinetic origin. The latter arises from the non-equilibrium correction to the Fermi distribution function $f_0$: $f(\epsilon)=f_0(\epsilon)+f_1(\epsilon)$ to be found from the kinetic equation
\be
\label{kinur}
\frac{2\gamma_\textrm{in}\rho(\epsilon)}{\hbar}f_1(\epsilon)=I_{\textrm{mw}}(\epsilon) ,
\ee
where $\rho(\epsilon)$
is the DOS in the superconductor normalized to its normal-state value [in terms of the spectral angle introduced in Sec.\ \ref{S:theory}, $\rho(\epsilon)=\Re\cos\theta^R(\epsilon)$], and $I_{\textrm{mw}}(\epsilon)$ is the collision integral for the interaction with the electromagnetic field \cite{mooij1981enhancement}. According to Eq.\ \eqref{kinur}, the correction $f_1(\epsilon)$ becomes singular in the absence of inelastic relaxation. That is why the first term in Eq.\ (\ref{OmegaA}) contains $\gamma_\textrm{in}$ in the denominator, whereas the limit $\gamma_\textrm{in}\to0$ is taken elsewhere. The specific dependence of $\rho(\epsilon)$ on $\Delta$ renders $G$ in Eq.\ \eqref{OmegaA} to be a non-analytic function of the order parameter.

In the limit \eqref{Eliash-cond}, the function $G$ has been evaluated exactly for $\Gamma=0$, relevant for the evaluation of the gap- and $T_c$ enhancement without the \emph{dc} supercurrent in Ref.\ \onlinecite{eckern1979stability}. It has also been estimated in the presence of the supercurrent ($\Gamma$ is determined by the current density) in Ref.\ \onlinecite{van1984inelastic}. We discuss both of these cases below.

\subsection{Gap enhancement}
\label{SS:Eliash-gap}

In the absence of a \emph{dc} supercurrent ($\Gamma=0$), the dynamic response of a superconductor is characterized by the function $G_0(\Delta/\hbar\omega)=G(\Delta/\hbar\omega,0)$ given by \cite{eckern1979stability}
\begin{equation}
G_{0}(u)
=
\begin{cases}
2\pi u\left( 1-u^{2}\right)^{-1/2}, & u<1/2 , \\
4 [ K +4u^2 ( \Pi - K ) ]/(2u+1), & u>1/2 ,
\end{cases}
\label{G0}
\end{equation}
where $K=K(k)$ and $\Pi=\Pi(a,k)$ denote complete elliptic integrals of the first and the third kinds \cite{elliptic-comment}, 
and 
\begin{equation}
a=\left(\frac{1}{2u+1}\right)^2,
\qquad
k=\left(\frac{2u-1}{2u+1}\right)^2.
\label{ak}%
\end{equation}
$G_0(u)$ is a positive-value function with a cusp at $u=1/2$ (corresponding to a
maximum $2\pi/\sqrt3$) and
the following asymptotes:
\begin{equation}
G_0(u)
=
\begin{cases}
2\pi u, & u\rightarrow0 , \\
2\ln(2.9u)/u, & u\rightarrow\infty .
\end{cases}
\label{G0ims}
\end{equation}

The value of $\Delta$ for given $\alpha$, $\gamma_\textrm{in}$, $\omega$ and $T$ should be obtained from solving Eqs.\ \eqref{GL} and \eqref{OmegaA} with $\Gamma=0$ and $G=G_0(\Delta/\hbar\omega)$.
Superconductivity is said to be enhanced if $\Delta(T)$ with irradiation exceeds its value $\Delta_0(T)$ in the absence of the microwave field, which happens provided $\omega>\omega_\textrm{min}(T)$.
According to Eq.\ \eqref{OmegaA},
$\omega_\textrm{min}(T)$ is bounded from below by $\omega_\textrm{min,min} = \sqrt3\gamma_\textrm{in}/\hbar$ [corresponding to $2\hbar\omega=\Delta_0(T)$].
%
Note however that the resulting minimal frequency $\omega_\textrm{min,min}$ does not obey the inequality \eqref{Eliash-cond} under which Eq.\ \eqref{OmegaA} was derived. That means that the Eliashberg theory can only estimate $\hbar\omega_\textrm{min,min}\sim\gamma_\textrm{in}$ but cannot predict the exact coefficient. A more precise criterion for the gap enhancement will be formulated in Sec.\ \ref{sec:res}.

\subsection{Critical current}
\label{SS:current}
In the presence of the supercurrent ($\Gamma\neq0$), the GL equation (\ref{GL}) for the order parameter should be supplemented by the expression for the current:
\begin{equation}
\label{js-1}
j_s/j_0
=
\sqrt{\frac{\Gamma}{2(k_BT_c)^3}}
\int d\epsilon \, W(\epsilon) \left[1-2f(\epsilon)\right] ,
\end{equation}
where $W(\epsilon)$ is a weight function, which becomes $W(\epsilon)=\pi\epsilon\delta(|\epsilon|-\Delta)$ for small pair breaking
(a more general expression is given in Sec.\ \ref{SS:pert}).
The supercurrent density is naturally measured in units of
\be
\label{j0}
j_0 = e\nu k_B T_c \sqrt{\frac{D k_B T_c}{\hbar}} ,
\ee
where $\nu$ is the DOS at the Fermi level per one spin projection.

The critical value of the current density corresponds to $\Gamma_c=4k_B(T_c-T)/{3\pi}$. In order to evaluate the function $G$ in the presence of a supercurrent, one has to consider the pair-breaking effect of the latter on the spectral functions of the superconductor.
The pair breaking leads to the smearing of the DOS $\rho(\epsilon)$ and the peak in the function $W(\epsilon)$ characterized by a width $w=(3/2)\Delta\left(\Gamma_c/\Delta\right)^{2/3}$ \cite{Maki-Parks,AGmagnetic}. As a result, in the limit $\hbar\omega\ll w\ll \Delta$ the logarithmic integration for $G$ is cut off by $w$ instead of $\hbar\omega$ and the enhancement function $G$ becomes
\be
\label{G-Son}
G\left(\frac{\Delta}{\hbar\omega}, \frac{\Gamma_c}{\Delta}\right)
=
\frac{2\hbar\omega}{\Delta}\ln\left(9.9\Delta/w\right)
\ee
[compare with the second line of Eq.\ (\ref{G0ims})].

Equations \eqref{GL}, \eqref{OmegaA}, \eqref{js-1} and \eqref{G-Son} were used in Ref.\ \onlinecite{van1984inelastic} to extract the inelastic scattering rate from experimental data on the enhancement of the critical current as a function of frequency.

\section{Theory for arbitrary temperatures}
\label{S:theory}

\subsection{Keldysh sigma model}

The response of a disordered superconductor to microwave irradiation can be described by the dynamic Usadel equation for the quasiclassical Keldysh Green's function $\check g$ supplemented by the self-consistency equation for the time-dependent order parameter \cite{larkin1986nonequilibrium,Kopnin-book}. This tedious procedure is simplified as long as the \emph{ac} component of the vector potential $\mathbf{A}_1(t)$ is small and can be treated as a perturbation on top of the steady state in the presence of a static $\mathbf{A}_0$. However even in that case calculations are quite lengthy due to a nonlinear and nonlocal-in-time constraint imposed on $\check g$. To treat the problem we find it convenient to use the language of the nonlinear Keldysh $\sigma$ model for superconducting systems \cite{FLS}. Though we need it only at the saddle-point level equivalent to the Usadel equation, we will benefit from the standard machinery for expanding in terms of $W$ modes (diffusons and cooperons).

The zero-dimensional Keldysh $\sigma$ model is formulated in terms of the order parameter $\check\Delta(t)$ and the matter field $Q_{tt'}$ which bares two time (or energy) arguments and acts in the tensor product of the Nambu and Keldysh spaces, with the Pauli matrices $\tau_i$ and $\sigma_i$, respectively. At the saddle point, $Q$ coincides with the quasiclassical Green's function $\check g$. In what follows we will consider time (or energy) arguments as usual matrix indices, with matrix multiplication implying convolution in the time (or energy) domain. The $Q$ matrix satisfies the nonlinear constraint $Q^2=1$. The $\sigma$-model action (which determines the weight $e^{iS/\hbar}$ in the functional integral) reads
\begin{equation}
\label{action}
S
= \frac{i\pi}{\delta} \mathop{\rm Tr}\left(
\Sigma Q
-\frac{\hbar D}{2} \hspace{0.5pt}
\check{\mathbf{a}}\tau_{3} Q \hspace{0.5pt} \check{\mathbf{a}} \tau_{3}Q
\right)
- \frac{4}{\lambda\delta} \Tr \Delta\Delta_\textrm{q},
\end{equation}
where $\delta=1/\nu V$ is the mean level spacing in the sample ($\nu$ is the DOS at the Fermi level per one spin projection, $V$ is the volume of the superconductor), $\lambda$ is the dimensionless Cooper coupling, and $\Sigma$ is given by
\be
\label{Sigma}
\Sigma = i\epsilon\tau_3 - \check\Delta\tau_1 - \frac{\gamma_\textrm{in}}{2} Q_\textrm{res} .
\ee
In Eqs.\ \eqref{action} and \eqref{Sigma} we introduce the following matrices in the Keldysh space:
\be
\check\Delta
=
\Delta \sigma_0 + \Delta_\textrm{q} \sigma_1
, \qquad
\check{\textbf{a}}
=
\textbf{a} \sigma_0 + \textbf{a}_\textrm{q} \sigma_1 ,
\ee
where $\Delta(t)$ and $\mathbf{a}(t) = e\mathbf{A}(t)/\hbar c$ are classical fields (observables), while $\Delta_\textrm{q}(t)$ and $\mathbf{a}_\textrm{q}(t)$ are their quantum counterparts (source fields).

Inelastic relaxation is modeled by tunneling to a normal reservoir described by the last term in Eq.\ \eqref{Sigma}, with $\gamma_\textrm{in}$ proportional to the tunnel conductance. The reservoir is assumed to be at equilibrium with the temperature $T$:
\be
Q_\textrm{res}
=
\begin{pmatrix} 1 & 2F_0 \\ 0 & -1 \end{pmatrix}_\textrm{K} \otimes \tau_3 ,
\ee
where $F_0$ is diagonal in the energy representation, with $F_0(\epsilon)=1-2f_0(\epsilon)=\tanh(\epsilon/2T)$ being the thermal distribution function. The collision integral in our model of inelastic relaxation is equivalent to the one used in the Eliahberg theory, see the LHS of Eq. (\ref{kinur}).

\emph{In the absence of irradiation}, the saddle-point solution in the superconductor is diagonal in the energy space, $Q_{\epsilon\epsilon'} = 2\pi\delta(\epsilon-\epsilon')Q(\epsilon)$, where $Q(\epsilon)$ can be written as
\be
\label{Qsaddle}
Q(\epsilon)
=
\begin{pmatrix} Q^R(\epsilon) & [Q^R(\epsilon)-Q^A(\epsilon)]F_0(\epsilon) \\ 0 & Q^A(\epsilon) \end{pmatrix}_\textrm{K} ,
\ee
with
\begin{subequations}
\begin{gather}
Q^R(\epsilon)
=
\begin{pmatrix}
\cos\theta^R(\epsilon) & \sin\theta^R(\epsilon) \\
\sin\theta^R(\epsilon) & - \cos\theta^R(\epsilon)
\end{pmatrix}_\textrm{N} ,
\\
Q^A(\epsilon)
=
-
\begin{pmatrix}
\cos\theta^A(\epsilon) & \sin\theta^A(\epsilon) \\
\sin\theta^A(\epsilon) & - \cos\theta^A(\epsilon)
\end{pmatrix}_\textrm{N} .
\end{gather}
\end{subequations}
The spectral angles obey the symmetry relations $\theta^{A}(\epsilon)=-\theta^{R}(-\epsilon)=-[\theta^{R}(\epsilon)]^{\ast}$ and can be found from the saddle point (Usadel) equation
\be
\label{Usadel}
\Delta\cos\theta^{R}(\epsilon)+i\epsilon^{R}\sin\theta^{R}(\epsilon
)-\Gamma\sin\theta^{R}(\epsilon)\cos\theta^{R}(\epsilon)=0 ,
\ee
where $\epsilon^{R,A}=\epsilon\pm i\gamma_\textrm{in}/2$ and the depairing energy $\Gamma$ defined in Eq.\ \eqref{Gamma} plays the role of the spin-flip rate $\hbar/\tau_s$ for magnetic impurities \cite{Maki-Parks,AGmagnetic}.
The equilibrium value of the order parameter should be obtained from the self-consistency equation [derivative of the action \eqref{action} with respect to $\Delta_\textrm{q}$]
\be
\label{SCE}
\Delta
=
\frac{\lambda}{2} \int d\epsilon \, F_0(\epsilon) \Im\sin\theta^R(\epsilon) .
\ee

\begin{figure*}
\centering
\includegraphics{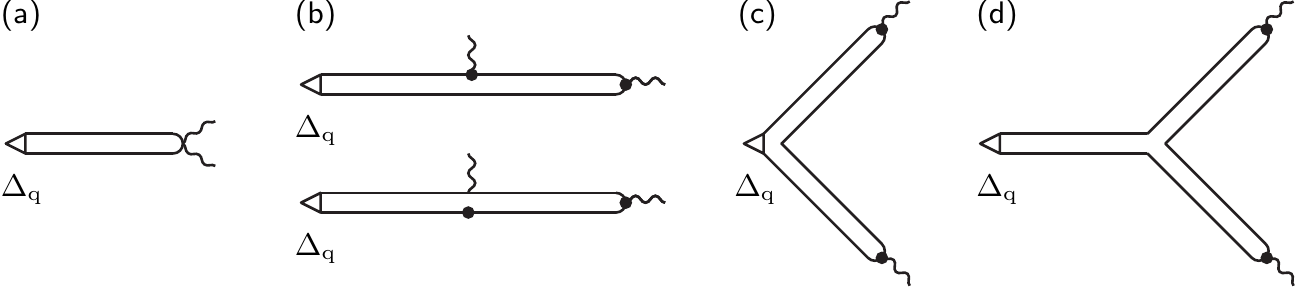}
\caption{Contributions to the action linear in the quantum conponent of the order parameter $\Delta_\textrm{q}$ and quadratic in $\mathbf{A}_1$ responsible for the term $F_\textrm{neq}(\Delta,\Gamma,T,\omega,\gamma_\textrm{in})$ in Eq.\ \eqref{SCE-gen}. Double lines stand for the diffusive modes \eqref{<WW>}, a triangle indicates $\Delta_\textrm{q}$, a wavy line denotes $\mathbf{A}_1\cos\omega t$, and a black dot denotes  $\mathbf{A}_0$. The difference between the diagrams (b) is that in the upper one each $W$ is extracted from its own $Q$, whereas in the lower one both $W$'s are extracted from the same $Q$.
The triple vertex in the diagram (d) involves the contribution from the cubic term in Eq.\ \eqref{Q-W}, which vanishes since the saddle satisfies the Usadel equation \eqref{Usadel}. The diagrams (b), (c) and (d) contain an additional factor of $A_0^2$ and therefore do not contribute to the modification of the gap in the absence of a \emph{dc} supercurrent.}
\label{fig:action}
\end{figure*}

\begin{figure*}
\centering
\includegraphics{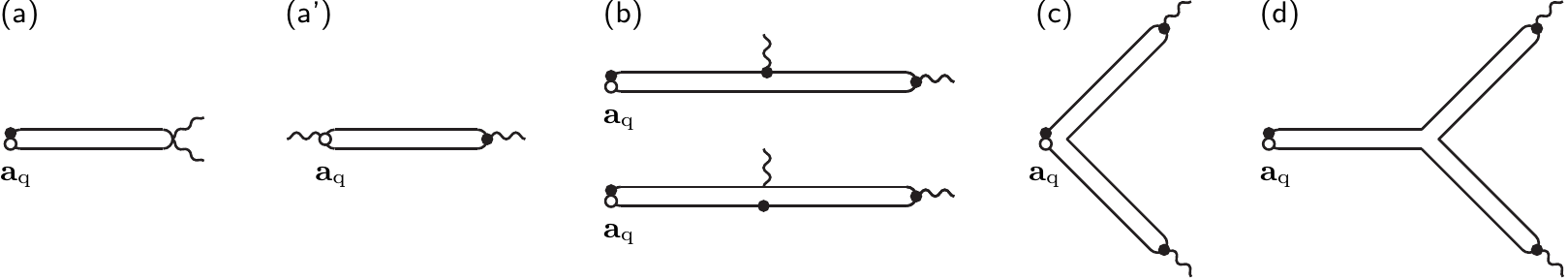}
\caption{Contributions to the action linear in the quantum component of the vector potential $\mathbf{a}_\textrm{q}$ and quadratic in $\mathbf{A}_1$ responsible for the term $I_\textrm{neq}(\Delta,\Gamma,T,\omega,\gamma_\textrm{in})$ in Eq.\ \eqref{j-gen}. The source field $\mathbf{a}_\textrm{q}$ is denoted by an open dot. The appearance of the diagram (a') which does not have its counterpart in Fig.\ \ref{fig:action} is due to quadratic coupling to the electromagnetic field in Eq.\ \eqref{action}.}
\label{fig:current}
\end{figure*}

\subsection{Diffusons and cooperons}

A microwave field $\mathbf{A}_1(t)$ drives the system out of equilibrium and induces non-diagonal in energy components of the matrix $Q$. In order to take them into account perturbatively, we parametrize small deviations from the saddle \eqref{Qsaddle} in terms of the matrix $W$ as \cite{KamenevAndreev,FLS,HouzetSkv,AntonenkoSkv}
\be
\label{Q-W}
Q=U_F^{-1}U^{-1}\sigma_3\tau_3 (1+W+W^2/2+\dots)UU_F ,
\ee
where the matrices $U$ and $U_F$ are diagonal in the energy representation:
\be
U_F
=
\begin{pmatrix} 1 & F \\ 0 & 1 \end{pmatrix}_\textrm{K} ,
\qquad
U
=
\begin{pmatrix} e^{i\tau_2\theta^R/2} & 0 \\ 0 & e^{i\tau_2\theta^A/2} \end{pmatrix}_\textrm{K} .
\ee
The parametrization \eqref{Q-W} reduces to the stationary saddle point \eqref{Qsaddle} at $W=0$ and automatically respects the nonlinear constraint $Q^2=1$ in the non-stationary case. Non-diagonal in energy elements of $Q$ are encoded by non-diagonal elements of $W$.

In general, a $4\times4$ matrix $W$ anticommuting with $\sigma_3\tau_3$ has eight nonzero elements. The \emph{ac} field $\mathbf{A}_1(t)$ excites only half of them that allows to restrict $W$ to the form
\be
W
=
\begin{pmatrix}
c^R i\tau_2 & d \tau_0 \\
-\overline d \tau_0 & c^A i\tau_2
\end{pmatrix}_\textrm{K} ,
\ee
where $c^R_{\epsilon\epsilon'}$ and $c^A_{\epsilon\epsilon'}$ are the cooperon modes responsible for the modification of the spectral angles $\theta^R$ and $\theta^A$, $d_{\epsilon\epsilon'}$ is the diffuson mode altering the distribution function, and $\overline{d}_{\epsilon\epsilon'}$ is its quantum counterpart.
The first-order correction to the spectral function is given by the following expression:
\begin{equation}
\label{dQR}
\delta Q^R_{\epsilon\epsilon'} = \left(\cos\frac{\theta^R_\epsilon+\theta^R_{\epsilon'}}{2}\tau_1-\sin\frac{\theta^R_\epsilon+\theta^R_{\epsilon'}}{2}\tau_3\right)c^R_{\epsilon\epsilon'}.
\end{equation}
The non-equilibrium correction to the distribution function is determined by $d_{\epsilon\epsilon'}$. Note that the upper right block 
of the matrix $W$ has only $\tau_0$ component. In the language of parametrization $Q^K=Q^R F - F Q^A$, conventional in the Usadel equation formalism, this implies $F$ being proportional to the identity matrix in the Nambu space. To the first order in $W$, one has
\begin{equation}
\label{dF}
\delta F_{\epsilon\epsilon'}
= \frac{d_{\epsilon\epsilon'}}{2\cos[(\theta^R_\eps-\theta^A_{\eps'})/2]}.
\end{equation}

Expanding the action \eqref{action} to the second order in $W$, we obtain the following bare correlation functions:
\begin{subequations}
\label{<WW>}
\begin{gather}
\label{C}
\corr{c^{R,A}_{\epsilon_1\epsilon_2}c^{R,A}_{\epsilon_3\epsilon_4}}
=
(\delta/\pi) \hat\delta_{\epsilon_1\epsilon_4} \hat\delta_{\epsilon_2\epsilon_3} C^{R,A}_{\epsilon_1\epsilon_2} , {}
\\
\label{D}
\corr{d_{\epsilon_1\epsilon_2}\overline d_{\epsilon_3\epsilon_4}}
=
(\delta/\pi) \hat\delta_{\epsilon_1\epsilon_4} \hat\delta_{\epsilon_2\epsilon_3} D_{\epsilon_1\epsilon_2} , {}
\end{gather}
\end{subequations}
where $\hat\delta_{\epsilon\epsilon'}=2\pi\delta(\epsilon-\epsilon')$ and the propagators of the diffusive modes are given by
\begin{subequations}
\label{CD}
\begin{gather}
C^\alpha_{\epsilon\epsilon'}
=
\frac{1}{\mathcal{E}^{\alpha\alpha}_{\epsilon\epsilon'}
+ \Gamma [1+\cos(\theta^\alpha_\epsilon-\theta^\alpha_{\epsilon'})] \cos(\theta^\alpha_\epsilon+\theta^\alpha_{\epsilon'})} ,
\\
D_{\epsilon\epsilon'}
=
\frac{1}{\mathcal{E}^{RA}_{\epsilon\epsilon'}
- \Gamma [1+\cos(\theta^R_\epsilon-\theta^A_{\epsilon'})] \cos(\theta^R_\epsilon+\theta^A_{\epsilon'})} .
\end{gather}
\end{subequations}
Here $\alpha=R, A$, and we use the notation $\mathcal{E}^{\alpha\beta}_{\epsilon\epsilon'} = \mathcal{E}^{\alpha}_{\epsilon} + \mathcal{E}^{\beta}_{\epsilon'}$ with
$
\mathcal{E}^{R,A}_{\epsilon}
=
\pm (-i\epsilon^{R,A}\cos\theta^{R,A}_\epsilon + \Delta\sin\theta^{R,A}_\epsilon)
$.

\begin{figure*}
\centering
\includegraphics{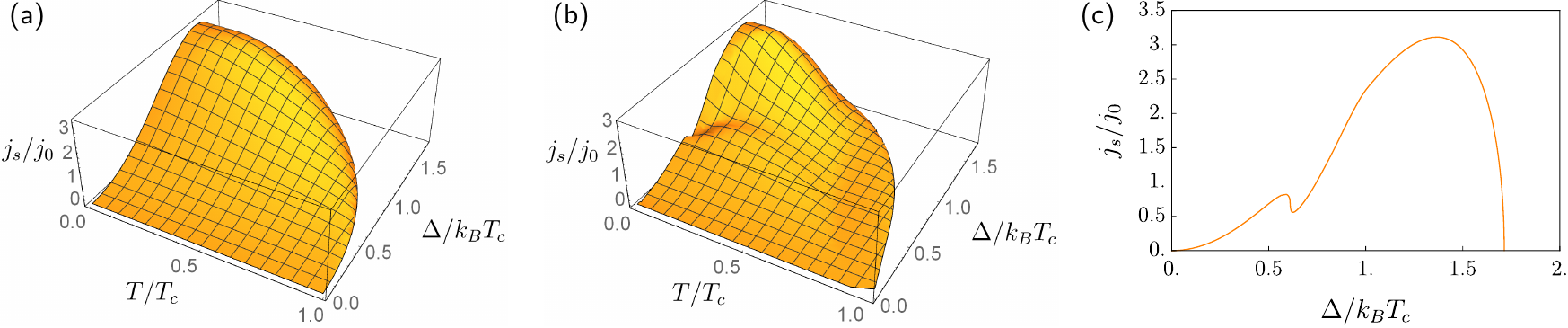}
\caption{(a) Critical current as a function of temperature and order parameter at equilibrium. 
(b)~Critical current under microwave irradiation with the frequency $\hbar\omega=0.1\,k_BT_c$, irradiation power $\alpha=0.1$ and inelastic scattering rate $\gamma_{in}=0.02\, k_BT_c$.
(c) Cross section of the surface (b) at $T=0$ showing an additional minimum in $j_s(\Delta)$ around $\Delta/k_BT_c\approx0.6$.}
\label{pd}
\end{figure*}

\subsection{Perturbative analysis of a microwave field}
\label{SS:pert}

In order to describe the full phase diagram of a superconductor at arbitrary temperatures and in the
presence of a \emph{dc} supercurrent, we need to generalize the GL equation (\ref{GL}) for arbitrary values of $\Delta$, $\Gamma$, $\omega$ and $T$. 

In the absence of microwaves, the equilibrium value of the order parameter $\Delta(\Gamma,T)$ should be obtained from a numerical solution of Eqs.\ \eqref{Usadel} and \eqref{SCE}. The supercurrent $j_s(\Gamma,T)$ is then calculated with the help of Eq.\ \eqref{js-1} with $1-2f(\epsilon)=F_0(\epsilon)$ and
$W(\epsilon)=2\Im\sin^2\theta^R(\epsilon)$, that leads to the critical current dependence $j_c(T)$ shown by the dashed line in Fig.\ \ref{jcall}.

In the presence of microwaves, the Usadel equation and the expression for the current are modified. The most effective way to study them is to consider the induced correction to the action.
In the second order in the magnitude of the \emph{ac} component of the vector potential \eqref{A}, we write it as
\begin{equation}
S[\Delta_\textrm{q},\mathbf{a}_\textrm{q}]
\approx
S_\textrm{eq}[\Delta_\textrm{q},\mathbf{a}_\textrm{q}]
+
\mathbf{A}_{1}^2
S_\textrm{neq}[\Delta_\textrm{q},\mathbf{a}_\textrm{q}]
,
\label{S+S}
\end{equation}
where $S_\textrm{eq}[\Delta_\textrm{q},\mathbf{a}_\textrm{q}]$ refers to the equilibrium case without irradiation. Here $\Delta_\textrm{q}$ and $\mathbf{a}_\textrm{q}$ are quantum sources needed to produce the self-consistency equation for the time-averaged order parameter $\Delta$ and the expression for the time-averaged supercurrent $j_s$ (in the absence of quantum sources, the action vanishes: $S[0,0]=0$).

The non-equilibrium correction to the action linear in $\Delta_\textrm{q}$ and quadratic in $\mathbf{A}_1$ is shown diagrammatically in Fig.\ \ref{fig:action}, where we keep only tree diagrams (no loops). The latter implies that we neglect quantum corrections and consider the saddle perturbed by a microwave field.
This formal scheme automatically takes into account corrections both to the spectral functions and the distribution function, since diffusive modes denoted by double lines in Fig.\ \ref{fig:action} can be either cooperons [Eq.\ \eqref{C}] or diffusons [Eq.\ \eqref{D}].
The resulting equation for the order parameter, $\delta S[\Delta_\textrm{q},0]/\delta\Delta_\textrm{q}|_{\Delta_\textrm{q}=0}=0$, can be written in the form
\begin{equation}
\mathcal{F}_\textrm{eq}(\Delta,\Gamma,T,\gamma_\textrm{in}) + \alpha \mathcal{F}_\textrm{neq}(\Delta,\Gamma,T,\omega,\gamma_\textrm{in}) = 0 ,
\label{SCE-gen}
\end{equation}
which can be considered as a generalization of the GL equation \eqref{GL} to the case of arbitrary temperatures.
To determine the supercurrent, one has to consider the non-equilibrium correction to the action linear in $\mathbf{a}_\textrm{q}$ and quadratic in $\mathbf{A}_1$ shown diagrammatically in Fig.\ \ref{fig:current}. Extracting the supercurrent density with the help of
$\mathbf{j}_s(t) = (ie/2V)\delta S[0,\mathbf{a}_\textrm{q}]/\delta\mathbf{a}_\textrm{q}(t)|_{\mathbf{a}_\textrm{q}=0}$ \cite{KamenevAndreev}, we get for the time-averaged supercurrent:
\begin{equation}
j_s/j_0
= \sqrt{\Gamma} [ \mathcal{I}_\textrm{eq}(\Delta,\Gamma,T,\gamma_\textrm{in}) + \alpha \mathcal{I}_\textrm{neq}(\Delta,\Gamma,T,\omega,\gamma_\textrm{in}) ] ,
\label{j-gen}
\end{equation}
where $j_0$ is defined in Eq.\ \eqref{j0}.

The key outcome of our theory are the functions $\mathcal{F}_\textrm{neq}(\Delta,\Gamma,T,\omega,\gamma_\textrm{in})$ and $\mathcal{I}_\textrm{neq}(\Delta,\Gamma,T,\omega,\gamma_\textrm{in})$. Simultaneous solution of Eqs.\ (\ref{SCE-gen}) and (\ref{j-gen}) gives the dependence of the order parameter $\Delta$ and the depairing rate $\Gamma$ on the temperature, \emph{dc} supercurrent, frequency and power of microwave irradiation, and the inelastic relaxation rate.

\section{Results}
\label{sec:res}

One of our results is presented in Fig.\ \ref{pd}(b), where the critical current under microwave irradiation is shown for $\alpha=0.1$, $\hbar\omega/k_BT_c=0.1$ and $\gamma_\text{in}/k_BT_c=0.02$. It is to be compared with the same dependence at equilibrium shown in Fig.\ \ref{pd}(a).
Remarkably, microwave irradiation strongly influences the phase diagram all over the parameter space. Two features can be clearly identified: (i) stimulated superconductivity in the vicinity of $T_c$ with Eliashberg-like enhancement [the lower right corner of Fig.\ \ref{pd}(b)], and (ii) strong sensitivity of the supercurrent to microwave radiation at low temperatures leading to the appearance of a pronounced minimum in $j_s(\Delta)$ around $\Delta/k_BT_c\approx0.6$ already for sufficiently weak driving power $\alpha$, see Fig.~\ref{pd}(c).

A complicated structure of the function $j_s(\Delta)$ at low temperatures with four solutions to the equation $j_s(\Delta)=j$ in a certain range of external currents $j$ raises the question of stability. At equilibrium, the stable branch with $dj_s/d\Delta<0$ is energetically favorable. Out of equilibrium, stability analysis becomes more involved \cite{schmid1977stability,eckern1979stability}. Note however that even if the non-equilibrium state with $\Delta\approx0.6\,k_B T_c$ is locally stable at low temperature, it might be very difficult to observe it experimentally. This question deserves future studies.

\subsection{Gap modification without supercurrent}
\label{SS:our-gap-enh}

While the general analysis of Eqs.\ (\ref{SCE-gen}) and (\ref{j-gen}) is rather complicated, one can derive the criterion for the gap enhancement in the absence of a \emph{dc} supercurrent ($\Gamma=0$). In this case, only the diagram shown in Fig.\ \ref{fig:action}(a) should be taken into account. Evaluating it and taking the derivative with respect to $\Delta_q$, we cast the resulting expression for the time-averaged order parameter in the form of Eq.\ \eqref{SCE-gen} with
\begin{equation}
\mathcal{F}_\textrm{eq}
=
\frac{1}{2\Delta} \int d\epsilon \,
F_0(\epsilon) \Im \sin\theta^R_\epsilon
- \frac{1}{\lambda}
\label{F-eq}
\end{equation}
and the non-equilibrium correction
\be 
\mathcal{F}_\textrm{neq} = \mathcal{F}_\textrm{neq}^\textrm{sp} + \mathcal{F}_\textrm{neq}^\textrm{kin}
\ee
being a sum of the spectral and kinetic contributions:
\begin{subequations}
\label{F&F}
\be
\label{F-neq-sp}
\mathcal{F}_\textrm{neq}^\textrm{sp}
=
- \frac{1}{4\Delta} \int d\epsilon \, F_0(\epsilon)
\Im \left\{ C^R_{\epsilon\epsilon}
\cos\theta^R_\epsilon \sin[\theta^R_\epsilon+\theta^R_{\epsilon-\omega}]
\right\} 
\ee
and
\begin{multline}
\label{F-neq-kin}
\mathcal{F}_\textrm{neq}^\textrm{kin}
=
\frac{1}{8\Delta} \int d\epsilon \, D_{\epsilon\epsilon} [F_0(\epsilon)-F_0(\epsilon-\omega)]
\\{}
\times \Im \left\{ \sin\theta^R_{\epsilon-\omega} -
\sin[\theta^R_{\epsilon-\omega}+\theta^R_\epsilon+\theta^A_\epsilon]
\right\} .
\end{multline}
\end{subequations}

\begin{figure}
\centering
\includegraphics[width=0.45\textwidth]{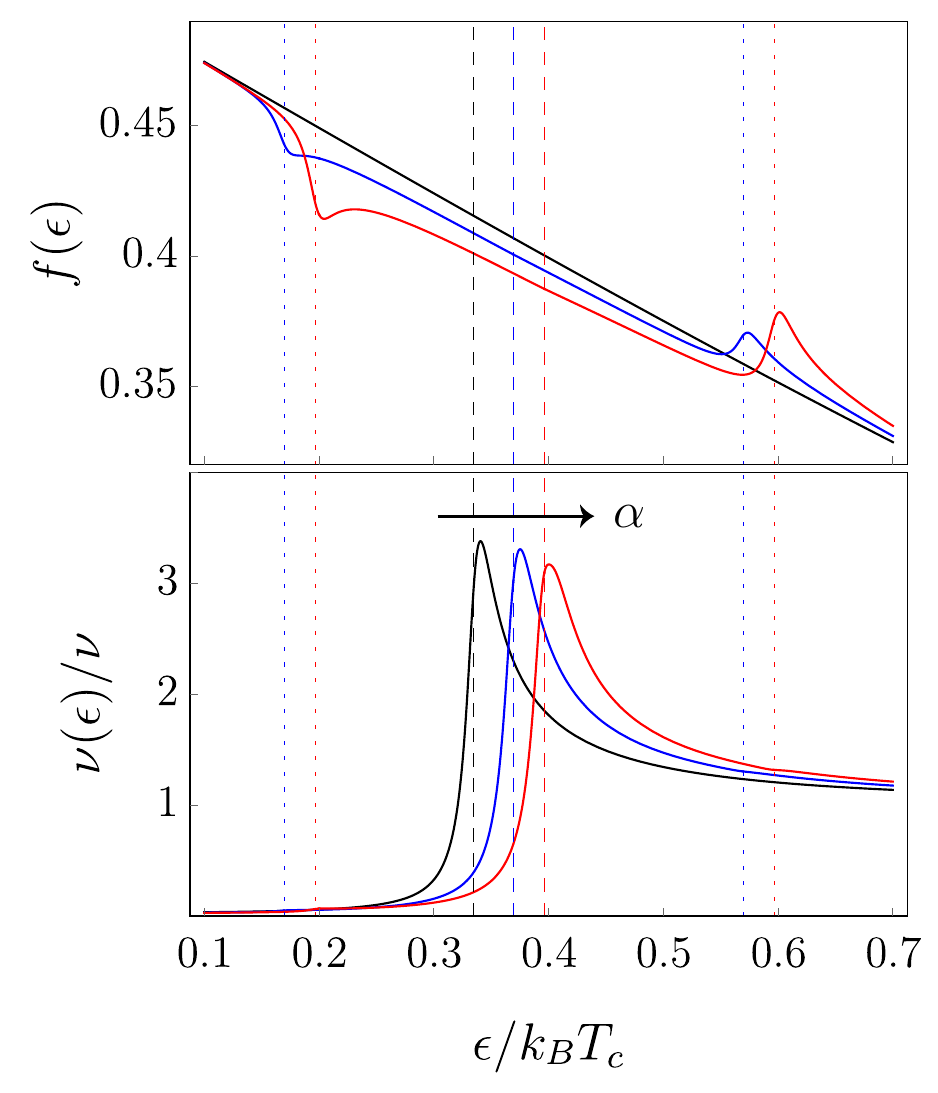}
\caption{Microwave-induced modification of the stationary (time-averaged) (a) quasiparticle distribution function $f(\eps)=[1-F(\eps)]/2$ and (b) density of states. 
Black to red: $\alpha/k_BT_c=0$, 0.005, and 0.01.
The curves are plotted at $T/T_c=0.98$, $\hbar\omega=20\gamma_{in}$, $\gamma_{in}/k_B T_c=0.02$, corresponding to the enhancement region in Fig. 1. The values of $\Delta$ marked by the vertical dashed lines are obtained from are obtained from the self-consistency equation.
Dotted lines correspond to $\Delta\pm\hbar\omega$.}
\label{dF-dnu}
\end{figure}

The results (\ref{F&F}) can be naturally interpreted as induced by the field-generated correction to the stationary (time-averaged) component of the spectral angle and the stationary (time-averaged) component of the distribution function,  correspondingly. Indeed, extracting the linear in $\alpha$ corrections to $\theta^R_\eps$ and $\delta F(\eps)$ from Eqs.\ (\ref{dQR}) and (\ref{dF}), we get
\begin{subequations}
\label{first-corr}
\be
\label{dthetaR1}
  \delta \theta^R_\eps
  = 
  - \frac{\alpha}{4}
  C^R_{\eps\eps}  
  \sin \left( \theta_\eps^R + \theta_{\eps-\omega}^R \right)
  + \{\omega\to-\omega\} 
\ee
and
\begin{multline}
\label{dF1}
  \delta F(\eps) 
  = 
  -
  \frac{\alpha D_{\eps\eps}[F(\eps)-F(\eps-\omega)]}{8\cos[(\theta^R_\eps-\theta^A_\eps)/2]} 
\\{}
  \times
  \left[ 
  \cos
   \left(\theta_{\eps-\omega}^R + \frac{\theta_\eps^R+\theta_\eps^A}{2} \right)
+ 
  \cos
   \left(\theta_{\eps-\omega}^A + \frac{\theta_\eps^R+\theta_\eps^A}{2} \right)
  \right]
\\{}
  + \{\omega\to-\omega\} .
\end{multline}
\end{subequations}
In Fig.\ \ref{dF-dnu}, we illustrate the influcence of microwaves on the stationary distribution function $f(E)=[1-F(E)]/2$ and the density of states $\nu(\eps)/\nu=\Re\cos\theta^R_\eps$.
Substituting now Eqs.\ (\ref{first-corr}) into the \emph{equilibrium}\/ expression (\ref{F-eq}), we recover the \emph{nonequilibrium}\/ contributions (\ref{F&F}).

We emphasize that spliting (\ref{F&F}) of $\mathcal{F}_\textrm{neq}$ into a sum of the spectral and kinetic contributions holds only in the absence of the supercurrent ($\mathbf{A}_0=0$). Then the \emph{ac}\/ component $\mathbf{A}_1(t)$ enters only squared, $\mathbf{A}_1^2(t)$, and only the diagram shown in Fig.\ \ref{fig:action}(a) contributes. This is not the case in the presense of the supercurrent, as the diagrams (b)--(d) suggest. This implies that in general, interpretation of the results in terms of time-averaged corrections to the distribution function and the spectral angle is impossible.

\subsubsection{\textbf{Comparison with the Eliashberg theory}}

Let us discuss how the Eliashberg theory is reproduced from Eqs.\ \eqref{F&F} at $T\to T_c$ in the limit \eqref{Eliash-cond}.
At equilibrium, equation $\mathcal{F}_\textrm{eq}=0$ coincides with the self-consistency equation \eqref{SCE}. In the vicinity of the transition, $-\mathcal{F}_\textrm{eq}$ gives the left-hand-side of the GL equation \eqref{GL} at $\Gamma=0$. The non-equilibrium terms then reproduce the right-hand side of Eq.\ \eqref{GL}.
Under the conditions \eqref{Eliash-cond}, the spectral contribution \eqref{F-neq-sp} gives $\mathcal{F}_\textrm{neq}^\textrm{sp} = -\pi/8k_BT_c$, reproducing the corresponding term in Eq.\ \eqref{OmegaA}. The kinetic contribution \eqref{F-neq-kin} contains the zero-frequency diffuson (loose diffuson \cite{KravtsovYudson}) $D_{\epsilon\epsilon}$, which is singular in the absence of inelastic relaxation
[compare with the kinetic equation \eqref{kinur}]. Keeping the leading order in $\gamma_\textrm{in}\to0$,
we find $\mathcal{F}_\textrm{neq}^\textrm{kin} = (\hbar\omega/16\gamma_\textrm{in}k_BT_c) G_0(\Delta/\hbar\omega)$, where $G_0(u)$ is given by Eq.\ \eqref{G0}. Hence we completely reproduce the main Eq.\ \eqref{OmegaA} of the Eliashberg theory in the limit \eqref{Eliash-cond}.

Our approach can be used to establish a refined criterion for the minimum frequency $\omega_\textrm{min,min}$ needed for the gap enhancement at some temperatures. As explained in Sec.~\ref{SS:Eliash-gap}, the simplified Eliashberg theory estimates $\hbar\omega_\textrm{min,min}\sim\gamma_\textrm{in}$ but fails to obtain the exact coefficient due to violation of the inequalities \eqref{Eliash-cond}.
On the other hand, our general equations \eqref{F&F} do not require those conditions to be fulfilled and can be applied for arbitrary $\omega/\gamma_\textrm{in}$.
In terms of the function $G_0(u)$, a finite value of $\omega/\gamma_\textrm{in}$ leads to the rounding of the cusp at $u=1/2$ and the overall suppression of the function. 
As a result, the enhancement effect becomes less pronounced and hence requires a larger frequency to be observable. We find
\be
\hbar\omega_\textrm{min,min} = 3.23 \, \gamma_\textrm{in},
\label{omega-min}
\ee
corresponding to $\hbar\omega_\textrm{min,min}/\Delta \approx 1.38$. 
This minimal frequency can be seen in the inset in Fig.~\ref{wmin-gap}. Equation \eqref{omega-min} is to be compared with the prediction of the simplified theory where the spectral smearing by $\gamma_\text{in}$ is neglected
[see Eq.\ \eqref{OmegaA}] that gives the factor $1.73$ instead of $3.23$ and the corresponding ratio $\hbar\omega_\textrm{min,min}/\Delta = 2$ \cite{mooij1981enhancement}.

\begin{figure}
\centering
\includegraphics[width=0.45\textwidth]{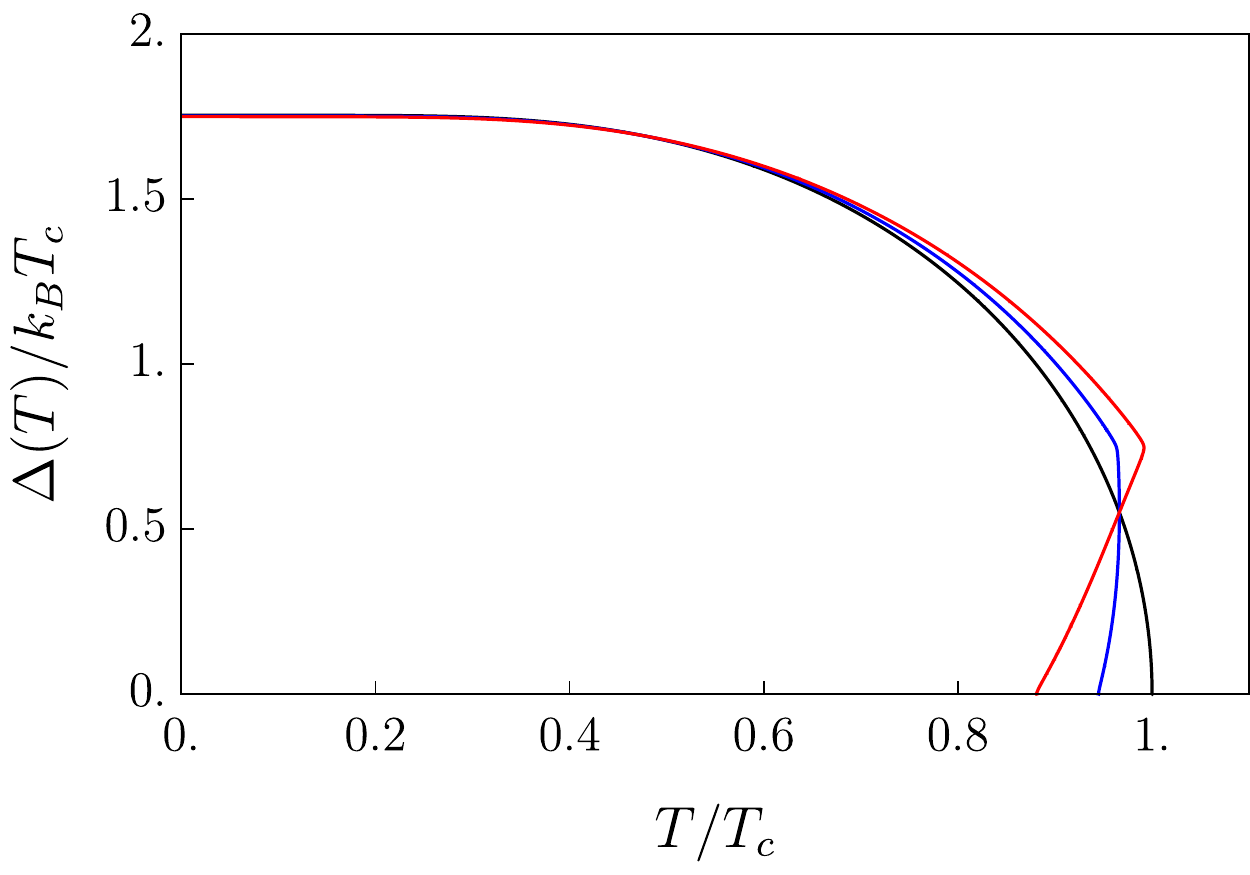}
\caption{Temperature dependence of the order parameter for zero \emph{dc} supercurrent.
Black: no microwaves (BCS case), color: microwave power $\alpha/k_BT_c = 0.005$ and $0.01$.
Microwave frequency $\hbar\omega/k_BT_c=1.5$, and $\gamma_\text{in}/k_BT_c=0.02$.
Gap enhancement near $T_c$ turns into gap suppression at low temperatures, in accordance with the phase diagram in Fig.\ \ref{wmin-gap}.}
\label{zero-current}
\end{figure}

\subsubsection{\textbf{Phase diagram at weak driving}}

The order parameter $\Delta(T)$ at given $\omega$, $\alpha$ and $\gamma_\text{in}$ should be obtained from a numerical solution of Eqs.\ \eqref{SCE-gen}, \eqref{F-eq}--\eqref{F&F}. 
To visualize the effect we compare the obtained $\Delta(T)$ with the equilibrium BCS value $\Delta_0(T)$ and identify the regions where the gap is enhanced [$\Delta(T)>\Delta_0(T)$] or suppressed [$\Delta(T)<\Delta_0(T)$]. 
A typical temperature dependence of the order parameter is shown in Fig.\ \ref{zero-current}. At some value of $\alpha>0$, the function $\Delta(T)$ becomes two-valued, with the upper (lower) branch being the stable (unstable) solution
\cite{schmid1977stability,eckern1979stability}.

The analysis simplifies in the limit of weak electromagnetic irradiation ($\alpha\to0$), where the boundary between the two regions is determined from the condition
\be
\label{boundary-eq}
\mathcal{F}_\textrm{neq}(\Delta_0(T),0,T,\omega,\gamma_\textrm{in}) = 0 
\ee
[the order of arguments as in Eq.\ \eqref{SCE-gen}].
For a given inelastic relaxation rate $\gamma_\text{in}$, the solution of this equation defines the curve $\mathcal{C}$ in the $(\omega,T)$ plane shown in Fig.~\ref{wmin-gap} for $\gamma_\text{in}/k_BT_c=0.02$. For small $\gamma_\text{in}$ this curve almost does not depend on $\gamma_\text{in}$, except for the vicinity of the critical temperature, where it marks the lower bound $\omega_\text{min,min}$ for the gap enhancement [see the inset to Fig.~\ref{wmin-gap} and Eq.\ \eqref{omega-min}].
Starting with $\omega_\text{min,min}$ near $T_c$, the lower part of the curve $\mathcal{C}$ describes the evolution of $\omega_\text{min}(T)$ with the temperature decrease. 

Remarkably, our results indicate that there exists also a maximal frequency $\omega_\text{max}(T)$ for gap enhancement. Thus the region of stimulated superconductivity encompassed by the curve $\mathcal{C}$ in Fig.~\ref{wmin-gap} is bounded both at low temperatures (no states available) and at high frequencies (heating-dominated regime).
A weak microwave signal cannot enhance $\Delta$ if the temperature is smaller than $T_\text{min}\approx 0.47 \, T_c$ or the frequency is larger than $\omega_\text{max}\approx 3.3 \, k_BT_c/\hbar$, despite of the fact that the distribution function continues to have a non-thermal structure. 

At small temperatures, $T\ll\Delta$, redistribution of quasiparticles (kinetic contribution) is not effective due to the suppressed DOS at low energies. Instead, the spectral contribution given by Eq.\ (\ref{F-neq-sp}) dominates. 
In the quasistationary limit, $\omega\ll\Delta$,
it turns to $\mathcal{F}_\textrm{neq}^\textrm{sp}=-\pi/8\Delta$.
At the same time, Eq.\ (\ref{SCE-gen}) becomes $\mathcal{F}_\textrm{eq}=\ln(\Delta/\Delta_0)$, and we get for the gap suppression: $\Delta=\Delta_0-\pi\alpha/8$. This is consistent with the Abrikosov-Gorkov result \cite{AGmagnetic,Semenov-2B} with the depairing rate $\alpha/2$ (the factor 1/2 is due to time averaging).

Finally, we would like to emphasize that the phase diagram shown in Fig.~\ref{wmin-gap} is plotted at vanishing microwave power, $\alpha\to0$. The main effect of small $\alpha$ is to shift the right boundary of the gap enhancement region to temperatures above $T_c$. Modification of the whole phase diagram as a function of $\alpha$ will be studied elsewhere \cite{we-2B}.

\begin{figure}[t]
\centering
\includegraphics[width=0.461\textwidth]{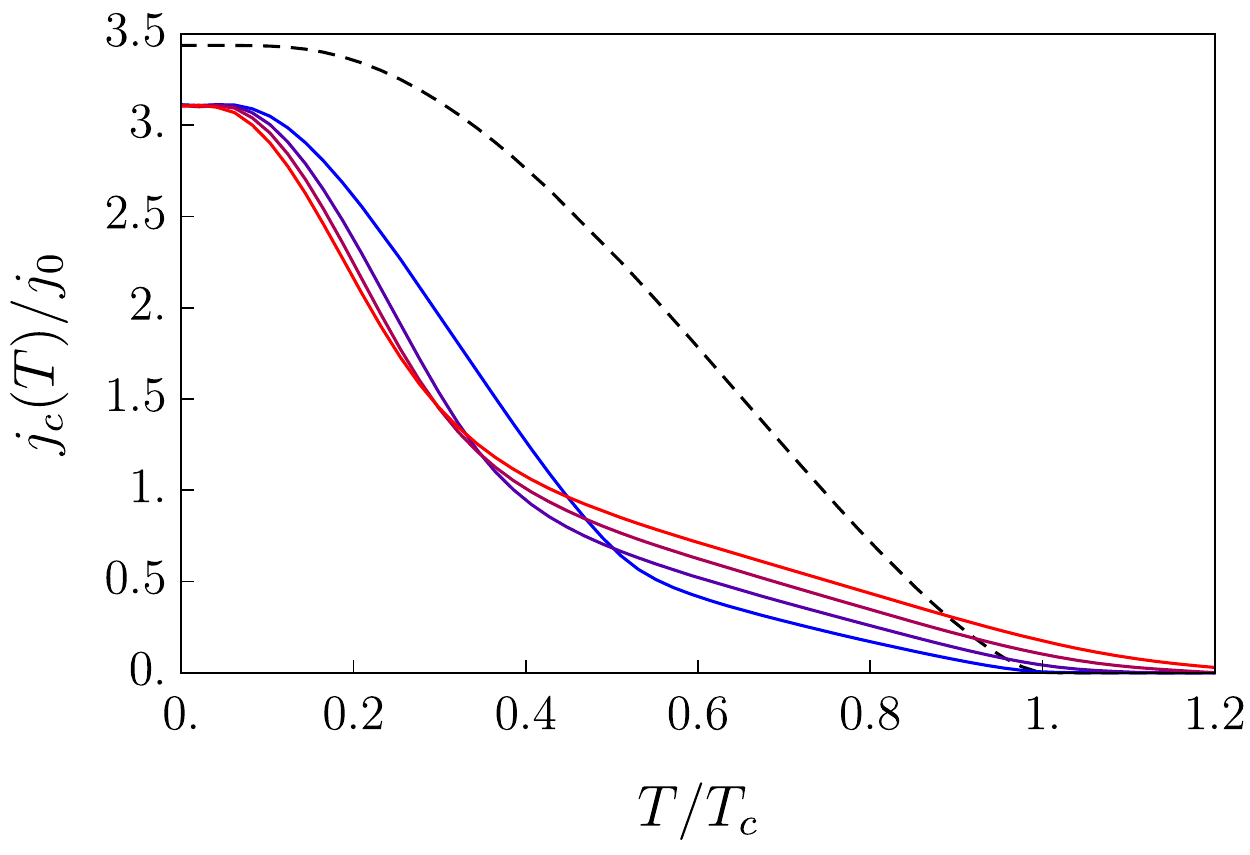}
\caption{Critical current as a function of temperature at fixed radiation intensity $\alpha=0.1\,k_B T_c$ and varying frequency. Black dashed line: without irradiation; color lines: frequency $0.1 \, k_B T_c/\hbar$ to $0.4 \, k_B T_c/\hbar$ from blue to red; $\gamma_\text{in}/k_BT_c=0.02$.}
\label{jcall}
\end{figure}

\subsection{Critical current enhancement}

Determination of the critical current $j_c(T)$ is a more complicated procedure, which requires maximization of the function $j_s(\Delta)$. In Fig.\ \ref{jcall} we plot the resulting $j_c(T)$ for a set of frequencies at a fixed irradiation power in the full range of temperatures. The dashed line is the critical current at equilibrium \cite{kupryanov1980temperature,romijn1982critical}. One clearly observes that the frequency, needed to enhance the supercurrent via irradiation at $T\sim T_c$, grows with the temperature decrease, consistent with previous studies. However at a certain $T$ of the order of $0.5\,T_c$, the sequence of the curves corresponding to various frequencies reverses. This happens when the effects of irradiation on the spectral properties of a superconductor (superconductivity suppression via pair-breaking) become more important than the kinetic effects (quasiparticle redistribution).

The region on the phase diagram where the critical current is enhanced by a weak microwave field is shown by the curve $\mathcal{C}'$ in Fig.~\ref{wmin-gap}. It is immersed into the region of gap enhancement enclosed by the curve $\mathcal{C}$, reflecting the fact that it is harder to stimulate superconductivity in the presence of depairing due to the supercurrent.

\section{Summary}
\label{sec:summary}

Using the formalism of the Keldysh nonlinear $\sigma$ model, we have studied the full phase diagram of a superconducting wire subject to the microwave irradiation in the presence of a \emph{dc} supercurrent. The only assumption is the small value of the amplitude of the \emph{ac} electromagnetic field, whereas all the other parameters of the theory can be arbitrary.
Our approach essentially generalizes the Eliashberg theory and the results for the critical current enhancement in the vicinity of $T_c$ \cite{schmid1977stability,eckern1979stability} to the case of arbitrary temperatures.
The developed theory treats the effect of quasiparticle redistribution on equal footing with the modification of the spectral properties.

One of our main findings is establishing the criteria for the microwave-stimulated enhancement (a) of the gap and (b) of the critical current, summarized in the phase diagram shown in Fig.\ \ref{wmin-gap}. We reveal that the gap enhancement is observed in a finite region of the $(\omega,T)$ plane, roughly limited by the conditions $T>0.5 \, T_c$ and $\hbar\omega<3 \, k_B T_c$. Such a behavior results from the interplay between several competing effects of the microwaves: 
(i) non-equilibrium distribution of quasiparticles with sub-thermal features 
responsible for stimulation of superconductivity, 
(ii) Joule heating,
and (iii) modification of the spectral functions due to depairing.
The absence of the gap enhancement at low $T$ should be attributed to the suppression of available quasiparticle DOS switching off the mechanism (i), whereas at large frequencies, the dominant effect is the Joule heating (ii).
In the presence of a supercurrent, the role of the mechanism (iii) is increased that makes the region of the critical current enhancement narrower than the region of the gap enhancement.

In our analysis we assumed the simplest model of inelastic relaxation by tunnel coupling to a normal reservoir. While its effect on the smearing of the BCS coherence peak is similar to that of electron-electron or electron-phonon interaction, it produces a notable DOS in the subgap region,
$\rho(\epsilon)=\Re[\epsilon^R/\sqrt{(\epsilon^R)^2-\Delta^2}]$, with an energy-independent Dynes-like parameter $\gamma_\textrm{in}/2$ \cite{Dynes}. As a result, the DOS is finite even at the Fermi level: $\rho(0)=\gamma_\textrm{in}/2\Delta\ll1$.
This suppresses the abovementioned mechanism (i) but does not turn it off since the left-hand side of the kinetic equation \eqref{kinur} remains finite in the limit $\gamma_\text{in}\to0$.
Therefore we expect that for a realistic energy-dependent $\gamma_\text{in}(\epsilon)$ the left boundary of the region of superconductivity enhancement in Fig.\ \ref{wmin-gap} may shift to higher temperatures.

Following the Eliashberg theory, our approach relies on the assumption of spatial homogeneity, when both the absolute value and the phase gradient of the order parameter are the same at every point in the wire. Then gauging out the phase one arrives at a zero-dimensional problem to be solved. Spontaneous breakdown of the translational symmetry leading to inhomogeneous non-equilibrium states was investigated in the framework of the Eliashberg theory in Ref.\ \cite{eckern1979stability}. It remains an open problem to study this effect for arbitrary temperatures.

The microwave response of superconductors at low temperatures has come into research focus recently \cite{de2014fluctuations,de2014evidence,sherman2015higgs,moor2017amplitude},
largely driven by applications of superconducting microresonators. For example, so called Microwave Kinetic Inductance Detectors (MKID) have been shown to be promising for astronomical studies \cite{Day2003,zmuidzinas2012superconducting,baselmans2017kilo}. 
In order to achieve a sufficiently high signal-to-noise ratio, given the existing low noise amplifiers, the microwave read-out signal is increased to a regime where a significant effect on the superconducting properties is observed. 
Our theoretical predictions can be used to analyze measurements on MKID \cite{de2014fluctuations,de2014evidence}, as well as in the experiment designed by Semenov \emph{et al}.\ \cite{Semenov-2B} (for application to a real experiment the nonlinear electrodynamics issues should be taken into account \cite{mooij1983nonlinear}). Apart from that, there are many controllable ways to drive superconducting systems out-of-equilibrium: disturbing them by a supercritical current pulse 
\cite{geier1982response,frank1983transient}, imposing to pulsed microwave phonons \cite{tredwell1975phonon},
or directly injecting non-equilibrium quasiparticles \cite{van1987enhancement1,van1987enhancement2}. It would be interesting to study these problems microscopically in the similar framework.

\acknowledgments

We are grateful to A. V. Semenov and I. A. Devyatov for stimulating discussions.
This research was partially supported by
the Russian Foundation for Basic Research (Grant No.\ 17-02-00757),
the Russian Science Foundation (Grant No.\ 17-72-30036),
and Skoltech NGP Program (Skoltech-MIT joint project).
TMK is also supported by the European Research Council Advanced
grant No.\ 339306 (METIQUM).

\bibliographystyle{apsrev}
\bibliography{scenh}

\begin{thebibliography}{51}
\bibitem[{ell()}]{elliptic-comment}
\bibinfo{note}{We use Wolfram Mathematica notations for elliptic integrals.}
\end{thebibliography}

\end{document}